\documentclass[12pt]{article}
\usepackage[english]{babel}
\usepackage{makeidx}
\usepackage{amssymb}
\usepackage{mathtools} 
\usepackage{mathrsfs}  
\usepackage{amsbsy}
\usepackage{amsmath}
\usepackage{times}
\usepackage{bm}
\usepackage{graphicx,color}
\usepackage{float}   
\usepackage{latexsym}
\usepackage{amssymb}
\usepackage{graphpap}
\usepackage{url}
\usepackage{indentfirst}
\usepackage{fancyhdr}
\usepackage{microtype}
\usepackage{authblk}

\usepackage{geometry} 
\date{} 

\title{Thermodynamics and Dynamic Stability: Extended Theories of Heat Conduction}

\author[1,4,5]{R. Somogyfoki}
\author[2]{A. Famà}
\author[2]{L. Restuccia}
\author[1,3,5]{P. Ván}
\affil[1]{Department of Theoretical Physics, Institute of Particle and Nuclear Physics, HUN-REN Wigner Research Centre for Physics, Konkoly Thege Miklós út. 29-33, Budapest H-1525, Hungary}
\affil[2]{Department of Mathematical and Computer Sciences, Physical Sciences and Earth Sciences, University of Messina, Messina, Italy}
\affil[3]{Department of Energy Engineering, Faculty of Mechanical Engineering, Budapest University of Technology and Economics,
	Műgyetem rkp. 3, Budapest 1111, Hungary}
\affil[4]{ELTE Doctoral School of Physics, Eötvös Loránd University, Pázmány Péter stny. 1/A, Budapest 1117, Hungary}
\affil[5]{Montavid Thermodynamic Research Group, Budapest, Hungary}

\begin{document}
	\maketitle
        \thispagestyle{empty}
	\abstract{The stability of homogeneous thermodynamic equilibrium is analyzed in heat conduction theories in the framework of nonequilibrium thermodynamics, where the internal energy, the heat flux and a second order tensor are thermodynamic state variables. It is shown, that the thermodynamic conditions of concave entropy and nonnegative entropy production can ensure the linear stability. Various special heat conduction theories, including Extended Thermodynamics, are compared in the general framework.}

	\section{Introduction}
	
	Thermodynamics and stability are closely connected. The tendency to homogeneous thermodynamic equilibrium in a neutral environment is our most general expectation regarding macroscopic stability of materials. According to the history of thermodynamics, this general experience was distilled in the Second Law \cite{Sen21b}. However, classical thermodynamics is a theory of homogeneous bodies, and, in spite of its name, is not a dynamical theory. Thermodynamic processes are different from mechanical ones: the classical theory of heat engines is not like the Newtonian theory of planet motion. Therefore, without evolution equations, the dynamic content of stability is a hidden aspect of the classical theory. The origin of this problematic situation is partially historical: the mathematical tools were missing when the thermodynamic ones were already clarified: the concept of Lyapunov stability appeared only in 1892 \cite{Lya892t}. At that time the Second and First Laws of thermodynamics were established knowledge in a formalism of equilibrium thermodynamics with a process concept but without time. Nevertheless, the stability of homogeneous thermodynamic equilibrium, that we will call {\em fundamental dynamic stability (FDS)}, remained a basic, but somehow implicit expectation in thermodynamic systems.   
		
	The spacetime formulation thermodynamics appeared only in 1940, in the framework of classical field theories when the Second Law of Thermodynamics was connected to the fundamental continuum balances with the calculation of entropy production. Classical Irreversible Thermodynamics, the classical field theory, was born at the same time for nonrelativistic and relativistic fluids \cite{Eck40a1,Eck40a3}. Classical multicomponent fluid systems of chemical thermodynamics profited the most from the uniform theory. Then the vision of the research field was the extension of the validity range considering memory and inertial effects, also in a relativistic framework, leading to theories of Extended Thermodynamics \cite{MacOns53a,Mul67a1,LebJou15a}. These theories introduced the dissipative fluxes of Classical Irreversible Thermodynamics as thermodynamical state variables \cite{Gya77a}. The primary benchmark was the compatibility with microscopic theories, first of all with kinetic theory of rarefied gases \cite{JouEta92b,MulRug98b}. The question of stability appeared in two contexts. First, the concept of dissipative structures based on the Second Law and on the entropy balance, \cite{GlaPri71b}; and also the mathematical conditions of extending the concepts of Lyapunov stability to inhomogeneous equilibrium solutions of nonlinear partial differential equations \cite{Gur75a,Sil97b}. However, it was clear from the beginning that dissipative structures appear in open systems, therefore the FDS of thermodynamic equilibrium is not an issue in this respect, it was and still is accepted as an evident intuitive consequence of the Second Law. On the other hand, mathematical investigations are struggling with the proper interpretation of the physical concepts, in particular the Second Law \cite{Daf79a,DosPru22a}. For example, considering the entropy production as a candidate of Lyapunov functional of homogeneous stationary states \cite{GlaPri71b}, is a misleading endeavour. 
	
	The analogy between the properties of entropy and the properties of a Lyapunov function is clear: entropy is concave and increasing along processes like a Lyapunov function. However, there is a twofold conceptual difficulty: for continuum theories the mathematics of Lyapunov stability is complicated, on the other hand, in thermodynamics of homogeneous bodies there is no time, there is no dynamics. Moreover, the attempts to understand the transition from continuum to homogeneous through various homogenisation methods encounter ill-defined process and equilibrium definitions \cite{GroMaz62b,TruBha77b,Mat92a1}.
	
	The formulation of thermodynamics of homogeneous bodies -- traditionally called thermostatics or equilibrium thermodynamics -- in a dynamical system framework was possible only with a proper definition of reservoir entropies and with the understanding that the simplest dynamics requires open systems, a body-reservoir connection. The modeling framework may be called {\em ordinary thermodynamics} because it reformulates the complete theory of Gibbsian equilibrium thermodynamics in a dynamical system framework and also because it is based on well-defined dynamical laws of ordinary differential equations \cite{Mat05b}.
	
	Ordinary thermodynamics is a theory of stability: the seemingly independent physical postulates are unified in a mathematically consistent stability framework with a clear physical meaning \cite{HadEta09b,Had19b,Van23a1}. Then a question emerges whether and how the continuum theory of nonequilibrium thermodynamics -- with partial differential equations -- is connected to the ordinary differential equations of ordinary thermodynamics? Whether the thermodynamic origin of the constitutive equations of a continuum theory would ensure the FDS or not. This question is particularly interesting in the light of the recent development of nonequilibrium thermodynamics, where the evolution equations of the thermodynamic state variables themselves are constructed with thermodynamic principles, like in Extended Thermodynamics and in case of Internal Variables \cite{MulRug98b,JouEta92b,MauMus94a1,BerVan17b}. It is particularly important to analyse the relation of universal theories (like gravity or quantum physics) and thermodynamic principles. The stability background of thermodynamics could be an explanation of their apparent universality \cite{Van23a}. 
	
	However, a too general approach may conceal the real problems. A simple necessary stability condition is provided by the dispersion relations: the conditions of the linear stability of homogeneous equilibrium can be easily calculated. It is far from being a pre-played game: there are several examples when FDS is expected, but not fulfilled. The most notable one is the theory of one component dissipative relativistic fluids, which is a major challenge of thermodynamics as a stability theory. It is proved that the homogeneous equilibrium of first-order dissipative relativistic fluids is unstable \cite{HisLin85a}\footnote{The classification of dissipative fluids is based on the moment series expansion of the parallel kinetic theory. Therefore for nonrelativistic generalised heat conduction first order is the Fourier theory, second order are the theories with heat flux $q^i$ and the presence of the flux of the heat flux $Q^{ij}$ indicates a third order theory. {\color{black} The classification terminology originates from relativistic theories \cite{HisLin85a,HisLin87a}.}}. The instability seemingly disappears in Landau--Lifshitz flow-frame, and according to detailed analysis it is clearly connected to the energy transport, to heat conduction \cite{VanBir08a}. The connection of thermodynamics and stability of homogeneous equilibrium for first-order dissipative relativistic fluids is an open problem. Recent investigations suggest that a special choice of flow-frames may ensure the expected FDS, at least in some special cases \cite{Kov19a,BeaFig23m}. The second order fluid theory of relativistic Extended Thermodynamics promised to restore the fundamental stability properties of relativistic fluids. However, the stability conditions of the homogeneous equilibrium of the Müller--Israel--Stewart theory, the simplest inertial extension of the first order one, seemingly does not follow from thermodynamic principles \cite{HisLin83a}, their interpretation in a thermodynamic framework is not straightforward \cite{GavEta20a}. For higher-order theories, there are no related direct investigations.
	
	In case of nonrelativistic dissipative fluids, the linear FDS is proved for the Fourier--Navier--Stokes system, and, according to the expectations, the conditions are purely thermodynamical \cite{van2009generic}\footnote{The stability is valid for any EOS with concave entropy density and with nonnegative viscosities and heat conduction coefficient. Unexpectedly, \cite{van2009generic} is the first general proof of this kind in the literature.}. The question arises, what are the conditions of FDS in the second- and higher-order theories of nonrelativistic continua? 
	
	\begin{figure}[h]
		\centering
		\includegraphics[width=0.8\textwidth]{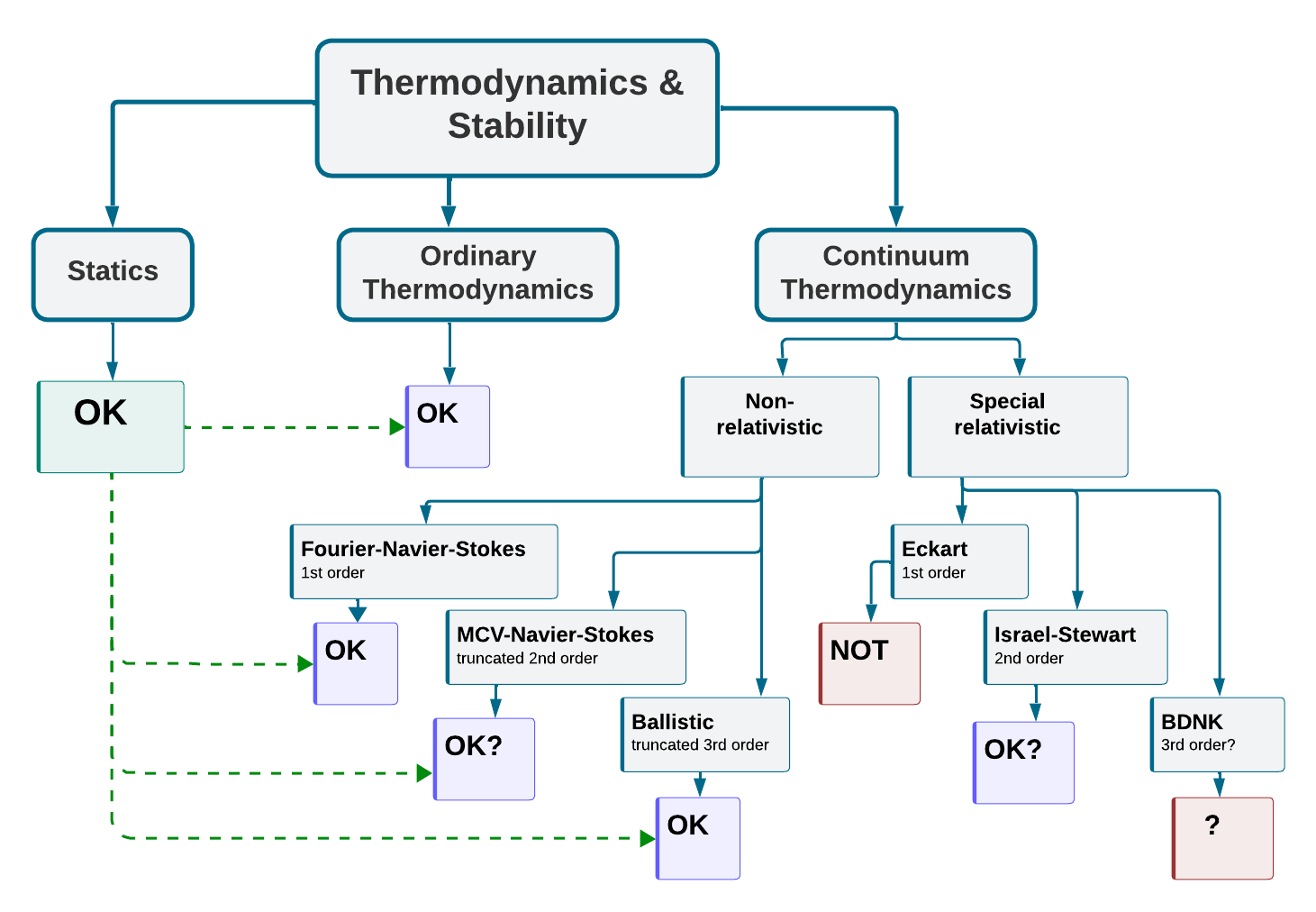}
		\caption{The hierarchy of nonrelativistic and special relativistic fluids and the role of thermodynamic conditions in the stability. For thermostatics it is only the concave entropy, but for ordinary thermodynamics it is Lyapunov stability. For continuum theories it is the linear stability of their homogeneous thermodynamic equilibrium. MCV stands for Maxwell--Cattaneo--Vernotte, and BDNK for Bemfica-- Disconzi--Noronha--Kovtun. \label{fig:thd_hierarchy}}
	\end{figure}
	
	In the following we investigate only heat conduction. It has several advantages: it is simpler, focuses to the problematic aspects and several heat conduction models are experimentally verified and tested. In general, non-Fourier heat conduction is a cornerstone and a test ground of nonequilibrium thermodynamics. Several models exist \cite{Kov24a}, and the various constitutive equations may have very different justifications. Thermodynamic compatibility sometimes is not a requirement and this is a source of unacceptable properties and instabilities \cite{KovVan18a1,Ruk14a,FabLaz14a}. Nonequilibrium thermodynamics with internal variables united most of the different models in a uniform framework, where the evolution equations for the heat flux (or a vectorial internal variable in general \cite{VanFul12a}) are derived with a constructive application of the Second Law of Thermodynamics \cite{KovVan15a}, and preserved the compatibility with Rational Extended Thermodynamics with notable differences \cite{KovEta21a}. The universality of the thermodynamic framework is justified by the theoretical prediction and experimental discovery of room temperature non-Fourier heat conduction in heterogeneous materials \cite{BotEta16a,VanEta17a}. In this framework the system of equations that lead to ballistic propagation is obtained, which is a third-order theory according to the terminology of relativistic fluid mechanics. 
	
	In this paper, the linear stability of homogeneous equilibrium of third order non-Fourier heat conduction in rigid heat conductors is analysed. The investigation is restricted to the most general nonequilibrium thermodynamic theory, which is directly compatible with kinetic theory of rarefied gases. Therefore, the basic fields are the temperature, the heat flux and a tensorial internal variable, and the corresponding theories will be called Extended Theories of Heat Conduction because the choice of the thermodynamic state variables is compatible with the ones Extended Thermodynamics. The general treatment with three spatial dimensions seems to be straightforward, but it is somehow complicated even with the simplest material symmetry in the isotropic case \cite{fama2021generalized}, therefore we restrict ourselves to one spatial dimension, where the corresponding generalisations can be easily classified and thermodynamic properties analysed.  
	
	The paper is organised as follows. In the next section the short outline of the nonequilibrium thermodynamic background of the system of equations is presented. Then the various forms of the constitutive evolution equations are treated in one spacial dimension. In section three the stability analysis is performed. In section four we treat various special theories and analyse the role of thermodynamic conditions and stability of homogeneous equilibrium. Our general conclusion is that thermodynamics ensures the linear stability of the homogeneous equilibrium, but, surprisingly, the particular choice of the internal variable as heat flux seems to be too restrictive. The discussion and some outlook are treated in the last section.
	
	\section{Third order Extended Heat Conduction -- an outline of thermodynamic background}
	
	
	In theories of heat conduction, the balance of internal energy is the starting point. It can be written as
	\begin{equation}\label{1}
		\rho  \dot e +  \partial_i q^i =0, 
	\end{equation}
	where the dot denotes the substantial (material) time derivative, $\dot e = \partial_t e+ v^i\partial_i e$, $\partial_i$ is the spatial derivative, $\rho$ is the mass density, $e$ is the specific internal energy, and $q^i$ is the heat flux, the conductive part of the current density of the internal energy, where $i=1,2,3$. Double indices denote the usual dual contraction, e.g. $\partial_i q^i =  \nabla\cdot \bf q$. 
	
	{\em Spacetime.} Equation \eqref{1} is the substantial form of the fundamental balance of energy conservation, where the spacetime derivative and the energy four-vector are split into timelike and spacelike parts in a frame that is comoving with the material \cite{Van17a}. The index notation is abstract; it has nothing to do with coordinates. It is also remarkable that the upper indices denote (contravariant) three-vectors and the lower ones three-covectors. The balance \eqref{1} is a four-divergence of the internal energy four-vector in the nonrelativistic spacetime.
	
	The material properties are determined by the following way:
	\begin{enumerate}
		\item {\em The thermodynamic state space and the constitutive functions}. Here the basic difference between theories is whether the heat flux is considered as a constitutive function or it is a basic field. Classical Irreversible Thermodynamics and Fourier theory  applies the first approach, in Extended Theories of Heat Conduction the second choice is unavoidable, if one keep compatibility with the moment series expansion of the Kinetic Theory \cite{CimEta14a,VanFul12a}. Here we will choose the latter one, therefore $q^i$ is considered as a basic field and the evolution equation of the heat flux will be determined constitutively. In this paper the thermodynamic state space is spanned by the specific internal energy $e$, the heat flux $q^i$, and by a second order three tensor field $Q^{ij}$, that turns out to be the flux of the heat flux. Naturally one requires evolution equations for both additional basic fields, $q_i$ and $Q^{ij}$, respectively.
		
		{\em Objectivity.} In this paper we deal with rigid heat conductors, therefore time derivatives are partial and density is constant, this way we can avoid most of the related complications like the connection of material time derivatives and well-posedness of the related mathematical problems \cite{ChrJor05a,GenStr24a}. According to the recent results of Angeles the natural choice of the Lie derivatives of the corresponding physical quantities is preferable \cite{Ang23a}. Thermodynamics may contribute to justify the proper choices and remove the seemingly ad-hoc use of various objective forms \cite{GioEta24m}. 
		
		\item  The second question is {\em the equation of state}, which determines the entropy density as a function of the thermodynamic state space variables. In the classical Fourier theory it is the caloric equation of state given as $e= c T$, where $T$ is the temperature and $c$ is the specific heat capacity. In our case it will be given in an extended form, where the entropy depends on the internal energy, the heat flux, $q^i$, and also on the the flux of the heat flux, $Q^{ij}$, as follows 
		\begin{equation}\label{stat_eos}
			s(e,q^i,Q^{ij}) =  s^{(eq)}(e) - \frac{1}{2\rho} q^i m_{ij} q^j - \frac{1}{2\rho} Q^{ij} M_{jilk}Q^{kl}.
		\end{equation}
		Here the specific entropy $s$ depends on the specific internal energy $e$ only in the equilibrium part $s^{(eq)}$, and the thermodynamic inductivities $m_{ij}$ and $M_{jilk}$ are assumed to be constants and positive definite. {\color{black} While internal variables can be seen as true dynamic degrees of freedom per Verhás \cite{verhas1997thermodynamics}, allowing a purely quadratic representation via the Morse lemma, this does not apply directly due to the fixed physical interpretations of heat flux, $q^i$, and of the flux of the heat flux, $Q^{ij}$. Further nonlinear terms like, $q^iQ_{ij}q^j$, may appear also in case of isotropic materials. The quadratic representation would align with Gaussian closures in the moment hierarchy \cite{Dre87a,Lev96a}. Hovewer, in the following linear stability analysis such nonlinear terms would not play any role.} The quadratic form ensures the simplest functional realisation of thermodynamic stability in all variables of the entropy function, together with a simple interpretation. Therefore $\beta = 1/T = \frac{d}{d e}s^{(eq)}$. If $s^{(eq)}$ is concave, that is $\frac{d^2 }{d e^2}s^{(eq)} \leq 0$, then the three-variable entropy function $s$ is concave, too, as it is seen calculating its negative definite second derivative:
			\begin{equation}\label{thds_gen}
				D^2 s = \begin{pmatrix}
					\frac{d^2 s^{(eq)}}{d e^2} = -\frac{1}{T^2}\frac{d T}{d e} & 0 & 0 \\
					0 & -m_{ij} & 0 \\
					0 & 0 & -M_{ijkl}
				\end{pmatrix}.
			\end{equation}
		\item The third important factor is {\em the entropy flux}, which is expected to be zero if $q^i$ and $Q{ij}$ are zero, and therefore it is best represented as:
		\begin{equation}\label{dyn_eos}
			J^i(e,q^i,Q^{ij}) = b^i_j q^j + B^i_{kj} Q^{jk}.
		\end{equation}
		Here the $b^i_j$ and $B^i_{kj}$ coefficient functions are the so-called Nyíri multipliers \cite{Nyi91a1}. 
	\end{enumerate}
	
	If the heat flux $q^i$ is considered a basic field, then one is looking for its evolution equation. The Nyíri multipliers are constitutive functions, that can depend on the basic fields and their gradients. They are to be calculated in the followings, using the methods of nonequilibrium thermodynamics.
	
	The constitutive functions determine the material properties, therefore they are restricted by the  Second Law of thermodynamics, the entropy balance:
	\begin{gather}\label{sbal}
		\rho \dot s + \partial_i J^i = \Sigma\geq 0.
	\end{gather}
	
	Considering \eqref{dyn_eos} and \eqref{stat_eos} one can calculate the entropy production rate and regroup the terms into the following convenient form:
	
	\begin{gather}\label{sprod}
		\Sigma = - q^i\left(m_{ij} \dot q^j- \partial_j b^j_i\right)+
		\left(b^i_j -\frac{1}{T}\delta^i_j\right)\partial_iq^j -
		Q^{ij}\left(M_{jilk}\dot Q^{kl} - \partial_kB^k_{ji}\right)+
		B^i_{kj}\partial_iQ^{jk} \geq 0.
	\end{gather}
	
	The Nyíri multipliers $b^i_j$ and $B^i_{kj}$ are not the only constitutive functions: the evolution equations, more properly, the right-hand side of the following formulae: $\dot q^i = f^{i}$ and $\dot Q^{ij} = F^{ij}$ are constitutive functions, too. Then the linear solution of the entropy inequality is the following:
	\begin{gather}\label{Onsgen}
		\begin{pmatrix}
			m_{ij} f^j- \partial_j b^j_i \\
			b^j_k -\frac{1}{T}\delta^j_k \\
			M_{lmvw} F^{wv} - \partial_wB^w_{lm}\\
			B^n_{op}
		\end{pmatrix} =
		\begin{pmatrix}
			^{11}L_{ia}^{} & ^{12}L_{ic}^{b} & ^{13}L_{ide}^{} & ^{14}L_{igh}^{f} \\
			^{21}L^j_{ka} & ^{22}L^{j\;\;b}_{\;kc} & ^{23}L^j_{kde} & ^{24}L^{j\;\;f}_{\;kgh} \\
			^{31}L_{lma}^{} & ^{32}L_{lmc}^{b} & ^{33}L_{lmde}^{} & ^{34}L_{lmgh}^{f} \\
			^{41}L^n_{opa} & ^{42}L^{n\;\;b}_{opc} & ^{43}L^n_{opde} & ^{44}L^{n\;\;\;f}_{opgh} 
		\end{pmatrix}
		\begin{pmatrix}
			-q^a\\
			\partial_bq^c \\
			-Q^{ed}\\
			\partial_fQ^{hg} 
		\end{pmatrix}.
	\end{gather}

	
	Here the symmetric part of the coefficient matrix is positive definite, therefore the linear solution transforms the entropy inequality into a positive definite quadratic form, where the transport matrices reflect the symmetries of the continuum material. The simplest isotropic symmetry is treated in detail in \cite{fama2021generalized}. It leads to unexpected results, due to the third order tensorial thermodynamic forces and fluxes. The cross-effects are unexpected in the sense that they contradict the heuristic Curie principle: the representation theorems of isotropic tensors show, that thermodynamic interactions with different tensorial order may be coupled in isotropic materials, too.
	
	\subsection{One-dimensional heat transport}
	
	In the following  we restrict ourselves to one spatial dimension in Cartesian coordinates, where
	\begin{gather}\label{note1d}
		B:= B^1_{11}, \quad Q := Q^{11}, \quad q:= q^1, \quad b:= b^1_1.
	\end{gather}
	Also, the thermodynamic inductivities are reduced to single scalars $m:= m_{11}$ and $M = M_{1111}$ in one dimension, independently of material symmetries. Also according to thermodynamic stability,
	\begin{equation}\label{thdst1}
		m>0, \qquad \text{and}\qquad M>0.
	\end{equation}
	
	In the one dimensional case the entropy flux \eqref{dyn_eos} and the entropy production \eqref{sprod} can be written respectively as 
	\begin{gather}\label{dyn_eos1d}
		J_1 = b q + B Q,
		\\\label{sprod1d}
		\Sigma_1 =  - q\left(m \dot q- \partial_x b\right)+
		\left(b -\frac{1}{T}\right)\partial_xq-
		Q\left(M\dot Q - \partial_xB\right)+
		B\partial_xQ \geq 0.
	\end{gather}
	Here $\partial_x$ indicates the spatial derivative with respect to $x$. The field quantities are assumed to change only along the direction of the $x$ coordinate.  
	
	The linear solution of the entropy inequality becomes
	\begin{gather}\label{o1}
		m \dot q -\partial_x b= - \lambda_{11} q +\lambda_{12} \partial_x Q, \\\label{o2}
		b-\frac{1}{T} = \kappa_{11} \partial_x q-\kappa_{12} Q, \\\label{o3}
		M \dot Q - \partial_x B = \kappa_{21} \partial_x q-\kappa_{22} Q,\\\label{o4}
		B =-\lambda_{21} q + \lambda_{22} \partial_xQ.
	\end{gather}
	This is the \textit{balance form} of the system of equations, because \eqref{o1} and \eqref{o3} are similar to balances where \eqref{o2} and \eqref{o3} define the corresponding $-b$ and $-B$ current densities. The $\lambda$ coefficients indicate cross effects between vectors and third order tensors in isotropic materials. That follows from the representation theorems and are a counterexample of the Curie principle \cite{fama2021generalized}. 	
	
	The system of constitutive equations is the same as (93)-(96) in \cite{fama2021generalized}, with the following notation of the isotropic conductivity coefficients:
	\begin{gather}\label{Onsgen1d}
		\begin{pmatrix}
			\lambda_{11} & 0 			& 0 			& \lambda_{12} 	\\
			0 			 & \kappa_{11}  & \kappa_{12} 	& 0				\\
			0			 & \kappa_{21}  & \kappa_{22}   & 				\\
			\lambda_{21} & 0 			& 0 			& 	\lambda_{22}
		\end{pmatrix}
		:= 
		\begin{pmatrix}
			L^{(1)} 	& 0 		& 0			& -L^{(1,4)} \\
			0			& L^{(2)}   & -L^{(2,3)} & 0\\
			0		 	& -L^{(3,2)} & L^{(3)} 	& 0\\
			-L^{(4,1)} 	& 0 		& 0 		& C^{(4)} 
		\end{pmatrix}.
	\end{gather}
	The new notation on the left hand side is introduced, because the calculations below and also the recent applications and models are one dimensional. The differences in the signs are consistent with the flux-force system in our paper (and mistaken in \cite{fama2021generalized}), however, the conditions of nonnegative entropy production are the same. It is also remarkable that in  \cite{fama2021generalized} the odd and even parity of $Q^{ii}$ were analyzed separately. However, here there are no further assumptions because the general approach will be helpful in the stability analysis. Time reversal symmetry, microscopic or macroscopic irreversibility \cite{Mei75a}, will be discussed in the last section. 
	
	
	Substituting the constitutive functions  \eqref{o1}-\eqref{o4}, one can obtain the one dimensional entropy flux \eqref{dyn_eos1d}, and entropy production \eqref{sprod1d}, as follows:
	\begin{gather}\label{dyn_eos1d_c}
		J_1 = \frac{q}{T} + \kappa_{11}\partial_x \left(\frac{q^2}{2}\right) + \lambda_{22}\partial_x \left(\frac{Q^2}{2}\right) - \left(\kappa_{12}+\lambda_{21}\right) q Q,
		\\\label{sprod1d_c}
		\Sigma_1 =  \lambda_{11} q^2 + \lambda_{22} (\partial_x Q)^2 +(\lambda_{12}+\lambda_{21}) (-q)\partial_x Q+
		\kappa_{11} (\partial_xq)^2 + \kappa_{22} Q^2 +\\ \nonumber (\kappa_{12}+\kappa_{21}) (-Q)\partial_x q
		\geq 0.
	\end{gather}
	
	One can see from the quadratic form of the entropy production, that the conductivity coefficients must fulfil the following inequalities:
	\begin{gather}\label{stcond1}
		\lambda_{11}\geq 0, \quad \lambda_{22}\geq 0, \quad \lambda_{11}\lambda_{22}- \frac{(\lambda_{12}+\lambda_{21})^2}{4} \geq 0, \\\label{stcond2}
		\kappa_{11}\geq 0, \quad \kappa_{22}\geq 0, \quad \kappa_{11}\kappa_{22}- \frac{(\kappa_{12}+\kappa_{21})^2}{4} \geq 0.
	\end{gather}
	
	Let us introduce the symmetric and antisymmetric crosscoefficients with the help of the following notation:
	\begin{gather}\label{asym}
		\hat\lambda:=\frac{\lambda_{12}+\lambda_{21}}{2}, \quad \check\lambda:=\frac{\lambda_{12}-\lambda_{21}}{2}, \quad
		\hat\kappa = \frac{\kappa_{12}+\kappa_{21}}{2}, \quad 
		\check\kappa =\frac{\kappa_{12}-\kappa_{21}}{2}.
	\end{gather}
	Then one obtains the inequalities:
	\begin{gather}\label{stcond1tot}
		\Lambda := \lambda_{11}\lambda_{22}-\lambda_{12}\lambda_{21} = \lambda_{11}\lambda_{22}-\hat\lambda^2 +\check\lambda^2\geq 0, \\\label{stcond2tot}
		K:= \kappa_{11}\kappa_{22}- \kappa_{12}\kappa_{21} = \kappa_{11}\kappa_{22}- \hat\kappa^2 + \check\kappa^2 \geq 0.
	\end{gather}
	Therefore, the antisymmetric parts of the matrices positively contribute to the determinant. 
	
	Eliminating $b$ and $B$ from the system \eqref{o1}-\eqref{o4} one can get the evolution equations of $q$ and $Q$, that we will call the {\em transport form} of the equations:
	\begin{align}\label{r1}
		m\dot{q} + \lambda_{11}q-\kappa_{11}\partial_{xx}q+(\kappa_{12}-\lambda_{12})\partial_x Q -\partial_x \beta &= 0,\\
		\label{r2}
		M\dot{Q} +\kappa_{22}Q-\lambda_{22}\partial_{xx}Q+(\lambda_{21}-\kappa_{21})\partial_x q &=0,
	\end{align}
	where $\beta = 1/T$ is the reciprocal temperature.
	
	Also, we have assumed, that the conductivity coefficients $\kappa_{11}$, $\kappa_{12}$, $\lambda_{22}$ and $\lambda_{21}$ are constants. One can obtain the {\em rate equation} of the heat flux $q$ by eliminating the internal variable $Q$ as well:
	\begin{gather}\label{qr}
		mM\ddot{q} + (m\kappa_{22}+M\lambda_{11})\dot{q} +\kappa_{22}\lambda_{11 } q -
		(m\lambda_{22}+M\kappa_{11})\partial_{xx}\dot{q} -\\\nonumber
		(\Lambda+ K + \kappa_{12}\lambda_{21}+ \kappa_{21}\lambda_{12})\partial_{xx}q +
		\kappa_{11}\lambda_{22} \partial_{xxxx}q = \\ \nonumber
		M\partial_{x}\dot\beta + \kappa_{22}\partial_x\beta-\lambda_{22}\partial_{xxx} \beta.
	\end{gather}
	
	{\bf If  $\lambda_{11} \neq 0$ and $\kappa_{22}\neq 0$}, then one can introduce a more insightful notation for the coefficients:
	\begin{gather}\label{phcoeff1}
		\tau_Q := \frac{M}{\kappa_{22}}, \quad
		\tau_q := \frac{m}{\lambda_{11}}, \quad
		l_Q := \sqrt{\frac{\lambda_{22}}{\kappa_{22}}}, \quad
		l_q :=  \sqrt{\frac{\kappa_{11}}{\lambda_{11}}}, \quad
		\lambda_{th} :=  \frac{1}{\lambda_{11}},
	\end{gather} 
	where $\tau_Q$ and $\tau_q$ are relaxation times, $l_Q$ and $l_q$ are characteristic lengths and $\lambda_{th}$ is the thermodynamic heat conduction coefficient. Then, the Fourier heat conduction coefficient can be written as $\lambda_{F}=\frac{1}{T^2\lambda_{th}}$. The two versions of heat conduction coefficients appear because the thermodynamic force of heat conduction is the gradient of the reciprocal temperature and not the gradient of the temperature. Furthermore, we introduce a third characteristic length as 
	\begin{gather}\label{phcoeff2}
	    l_{qmod} :=  \sqrt{l_q^2 + a_qa_Q}, \quad \text{where} \quad
		a_q := \frac{\kappa_{12}-\lambda_{12}}{\lambda_{11}}, \quad  \text{and} \quad
		a_Q := \frac{\lambda_{21}-\kappa_{21}}{\kappa_{22}}.
	\end{gather}
	
	Therefore, the \textit{transport form} of the system of equations, \eqref{r1}-\eqref{r2}, can be written  with the new coefficients as
	\begin{align}\label{r1p}
		\tau_q\dot{q} + q-l_q^2\partial_{xx}q +a_Q\partial_x Q &=\lambda_{th}\partial_x \beta ,\\
		\label{r2p}
		\tau_Q\dot{Q} +Q-l_Q^2\partial_{xx}Q+a_q\partial_x q &=0.
	\end{align}

	Also, the \textit{rate equation} of the heat flux \eqref{qr} can be rewritten in the following form:
	\begin{gather}\label{qr1}
		\tau_Q\tau_q\ddot{q} + 	(\tau_q+\tau_Q)\dot{q} + q -
		(\tau_q l_Q^2 +\tau_Ql_q^2)\partial_{xx}\dot{q} -
		(l_Q^2 + l_{qmod}^2)\partial_{xx}q + l_Q^2l_q^2 \partial_{xxxx}q - \\ \nonumber
		\lambda_{th} \left(\tau_Q\partial_{x}\dot\beta + \partial_x\beta-l_Q^2\partial_{xxx} \beta \right)=0.
	\end{gather}
	We can see that the number of relevant coefficients is reduced to two relaxation times and three characteristic length scales beyond the heat conduction coefficient. It is also important that the rate equation of the heat flux can be reorganised into the following instructive form:
	\begin{gather}\label{qr2}
		(\tau_Qd_t- l_Q^2\partial_{xx})(\tau_q\dot{q} + q -l_q^2\partial_{xx}\dot{q} - \lambda_{th}\partial_x\beta)+
		(\tau_q\dot{q} + q -l_{qmod}^2\partial_{xx}\dot{q} - \lambda_{th}\partial_x\beta)=0.
	\end{gather}
	Here $d_t$ appears as the time derivative operator ($d_t q= \dot q$). This is the {\it hierarchical form of the rate equation of the heat flux}. The hierarchical form is instructive, because it reflects the structure of the system of evolution equations before the elimination of the second order tensor component $Q$. Presumably, the above hierarchy is valid also in three (or two) spatial dimensions and characterizes the constraint of higher order heat conduction originated in the Second Law of Thermodynamics.
	
	Finally, the balance of internal energy \eqref{1}, can be written in one spatial dimension as
	\begin{equation}\label{eba_1d}
		\rho c_v\dot T + \partial_x q = 0, 
	\end{equation}
	where 
	\begin{equation}\label{thdst0}
		c_v = \frac{d e}{d T} > 0,
	\end{equation} 
	according to thermodynamic stability. This last equation completes the set of constitutive equations.

	\section{Exponential plane waves}\label{sec_3}
	
	In this section, analogously as in \cite{van2009generic}, we consider the  perturbations of the physical quantities  $T,q$ and $Q$  starting from their equilibrium values $T_0\neq 0$, $q_0=0$ and $Q_0=0$, respectively. Therefore, we have
	\begin{equation}\label{6}
		T\approx T_0+\delta T, \qquad q\approx \delta q, \qquad Q\approx \delta Q. 
	\end{equation}
	Then, from  the system  of perturbed equations
	\eqref{eba_1d}, \eqref{r1} and \eqref{r2} we obtain 
	\begin{align}
		\label{7}
		&\rho c_v\partial_t \delta T+\partial_x \delta q = 0, \\
		\label{8}
		&m \partial_t \delta q+\lambda_{11}\delta q-\kappa_{11}\partial_{xx}\delta q + 
		\frac{1}{T_0^2}\partial_x \delta T + (\kappa_{12} - \lambda_{12})\partial_x \delta Q = 0,\\
		\label{9}
		&M\partial_t \delta Q+\kappa_{22}\delta Q-\lambda_{22}\partial_{xx}\delta Q+(\lambda_{21}-\kappa_{21})\partial_x \delta q =0.
	\end{align}
	We  look for  the exponential plane wave solutions of this system of equations, i.e. 
	\begin{equation}
		\label{10}
		\delta T=\widehat{\delta T}e^{\Gamma t+ikx}, \qquad \delta q=\widehat{\delta q}e^{\Gamma t+ikx}, \qquad \delta Q=\widehat{\delta Q}e^{\Gamma t+ikx}.
	\end{equation}
	From these we obtain 
	\begin{align}
		\label{11}
		&\rho c_v\Gamma{\delta T} + ik{\delta q}=0, \\
		\label{12}
		&ik{\delta T}+T^2_0\left(m\Gamma+\lambda_{11}+\kappa_{11}k^2\right){\delta q} + T^2_0(\kappa_{12} - \lambda_{12})ik{\delta Q}=0, \\
		\label{13}
		&(\lambda_{21}-\kappa_{21})ik{\delta q} + \left(M\Gamma + \kappa_{22} + \lambda_{22}k^2\right){\delta Q}=0.
	\end{align} 
	The system  of equations \eqref{11}-\eqref{13} has non-trivial solutions only if its determinant vanishes, i.e.
	\begin{equation}
		\label{14}
		\begin{vmatrix}
			\rho c_v\Gamma & ik & 0 \\[0.5em]
			ik/T^2_0 & \left(m\Gamma+\lambda_{11}+\kappa_{11}k^2\right) & (\kappa_{12} - \lambda_{12})ik \\[0.5em]
			0 & (\lambda_{21}-\kappa_{21})ik & \left(M\Gamma+\kappa_{22}+\lambda_{22}k^2\right)
		\end{vmatrix}
		=0.
	\end{equation}
	From \eqref{14} we find the following polynomial:
	\begin{equation}\label{15}
		\begin{split}
			0& =mM\Gamma^3 + 
			\left[m\kappa_{22}+ M\lambda_{11}+ \left(m\lambda_{22} +M\kappa_{11}\right)k^2\right]\Gamma^2 +\\
			& +\left[\frac{k^2 M}{T^2_0 \rho c_v} + 
			\lambda_{11}\kappa_{22} +
			\left(\Lambda+K+ \lambda_{12}\kappa_{21}+\lambda_{21}\kappa_{12} \right)k^2 + 
			\lambda_{22}\kappa_{11}k^4\right]\Gamma +\\
			& +\frac{k^2}{T^2_0 \rho c_v}\left(\kappa_{22}+\lambda_{22}k^2\right).
		\end{split}
	\end{equation}
	
	If $\lambda_{11}\neq 0$ and $\kappa_{22}\neq 0$, then we can rewrite the stability polynomial with the physical parameters \eqref{phcoeff1} and \eqref{phcoeff2} and obtain
	\begin{equation}\label{15b}
		\begin{split}
			0& =\tau_q\tau_A\Gamma^3 + 
			\left[\tau_q+ \tau_Q+ \left(\tau_ql_{Q}^2 +\tau_Ql_q^2\right)k^2\right]\Gamma^2 +\\
			& +\left[1 + 
			\left(\tau_Q\alpha + l_q^2 + l_Q^2 + a_q a_Q \right)k^2 + 
			l_q^2 l_Q^2 k^4\right]\Gamma + k^2\alpha\left(1 + l_Q^2k^2\right).
		\end{split}
	\end{equation}
	
	where $\alpha = \frac{1}{\lambda_11 T_0^2 c_v\rho}= \frac{\lambda_F}{c_v\rho}$ is the thermal diffusivity. 
	
	To find the conditions of  stability for   the exponential plane waves \eqref{10} we discuss the \eqref{15} dispersion relation. It has  the form
	\begin{equation}
		\label{16}
		a_0\Gamma^3+a_1\Gamma^2+a_2\Gamma+a_3=0,
	\end{equation}
	where   $a_0, a_1, a_2$ and $ a_3$  are  in equation \eqref{10}, respectively, the coefficients of $\Gamma^3, \Gamma^2, \Gamma$, and the last term $\Gamma^0$.  
	Then, according to the Routh-Hurwitz criteria (see \cite{KorKor00b}), the real parts of the solutions of \eqref{16} are negative if every coefficient is nonnegative and also
	\begin{align}\label{17}
		0 & < a_1a_2-a_0a_3 = \\\nonumber
		& = \kappa_{11}\lambda_{22} \left(m\lambda_{22}+M\kappa_{11}\right)k^6+ \\\nonumber
		& + \left[\frac{M^2\kappa_{11}}{\rho c_vT^2_0} + \kappa_{11}\lambda_{22} \left(m\kappa_{22}+M\lambda_{11}\right)+ \right. \\\nonumber
		& \qquad \left. +\left(m\lambda_{22}+M\kappa_{11}\right)\left(\underline{\Lambda+K+\lambda_{12}\kappa_{21}+\lambda_{21}\kappa_{12}}\right)\right]k^4 + \\\nonumber
		& + \left[\frac{M^2\lambda_{11}}{\rho c_vT^2_0} + \lambda_{11}\kappa_{22} \left(m\lambda_{22}+M\kappa_{11}\right)+ \right. \\\nonumber
		& \qquad \left. +\left(m\kappa_{22}+M\lambda_{11}\right)\left(\underline{\Lambda+K+\lambda_{12}\kappa_{21}+\lambda_{21}\kappa_{12}}\right)\right]k^2 + \\\nonumber
		& +\lambda_{11}\kappa_{22}\left(m\kappa_{22}+M\lambda_{11}\right),
	\end{align}
	
	or, expressed with the physical parameters 
	\begin{align}\label{17a}
		0 & < a_1a_2-a_0a_3 = \\\nonumber
		& = l_q^2l_Q^2 \left(\tau_ql_Q^2+\tau_Ql_q^2\right)k^6+ \\\nonumber
		& + \left[\tau_Q^2l_q^2\alpha + l_q^2l_Q^2 \left(\tau_q+\tau_Q\right) +
		\left(\tau_ql_Q^2+\tau_Ql_q^2\right)\left(\underline{l_q^2 + l_Q^2 + a_q a_Q }\right)\right]k^4 + \\\nonumber
		& + \left[\tau_Q^2\alpha + \left(\tau_q l_Q^2+\tau_Q l_q^2\right)
		+\left(\tau_q+\tau_Q\right)\left(\underline{l_q^2 + l_Q^2 + a_q a_Q }\right)\right]k^2 + \\\nonumber
		& +\left(\tau_q+\tau_Q\right)
	\end{align}
	
	One can see, that $a_0>0, a_1>0, a_3>0$, follow from pure thermodynamic conditions, namely from thermodynamic stability \eqref{thdst0}, \eqref{thdst1} and the inequalities, \eqref{stcond1} and \eqref{stcond2}, required by nonnegative entropy production. However, the inequalities \eqref{17} and \eqref{17a} and the condition $a_2>0$ are fulfilled only if 
	\begin{gather}\label{dstabcond}
		\Lambda+K+ \lambda_{12}\kappa_{21}+\lambda_{21}\kappa_{12} = \\\nonumber
		\lambda_{11}\lambda_{22} + \kappa_{11}\kappa_{22} - (\hat\lambda-\hat\kappa)^2 + (\check\lambda-\check\kappa)^2\geq 0,
	\end{gather}
	which, expressed with the physical parameters becomes
	\begin{gather}\label{dstabcond2}
		l_q^2+l_Q^2+a_qa_Q\geq 0.
	\end{gather}
	
	Here we have introduced the notation stated in equation \eqref{asym}, for the symmetric and antisymmetric crosscoefficients. It is apparent from the second form of the inequality that the thermodynamic conditions for the conductivity coefficients \eqref{stcond1tot}-\eqref{stcond2tot} ensure only that 
	$\lambda_{11}\lambda_{22} -\hat\lambda^2 + \hat\kappa_{11}\kappa_{22} -\kappa^2 + (\check\lambda-\check\kappa)^2\geq 0$
	and the possibility 
	\begin{gather}\label{dstabcond1}
		\lambda_{11}\lambda_{22} -\hat\lambda^2 + \kappa_{11}\kappa_{22} -\hat\kappa^2 + (\check\lambda-\check\kappa)^2\leq -2\hat\lambda\hat\kappa 
	\end{gather}
	does not violate the second law. Interestingly the violation requires the symmetric parts of the cross-coefficients to be $\hat\lambda\neq  0$ and $\hat\kappa\neq 0$. Those parameters contribute to the dissipative part of the entropy production. If $\hat\lambda$ or $\hat\kappa$, are zero, then the stability condition \eqref{dstabcond2} is satisfied. In the known, hyperbolic framework of Extended Thermodynamics these parameters are zero, therefore the theory is stable, as we will see in the next section, where the special cases of the general framework will be analysed.

    \section{Special cases}
	
	The special cases are best analysed with conditions for the original thermodynamic coefficients, because those represent the proper material properties and the most general form of the second law restrictions. However, the physical parameters provide a simpler interpretation and insight, therefore we first impose the conditions in physical parameters, then extend the analysis to the cases that are not tractable with them. 

    The analysis is based on the physical insight that from the experimental point of view only those special heat conduction equations are viable that are obtained in the absence of a particular process whose existence is determined by properties of materials. One cannot eliminate a term in the rate equation \eqref{qr} arbitrarily, only conditions for the thermodynamic parameters have physical relevance. 
	
	\subsection{Decoupled heat conduction: $a_q=0$ or $a_Q=0$}
	
	It is easy to see, that the previous general system of constitutive partial differential equations \eqref{r1p}-\eqref{r2p} becomes decoupled if either $a_q=0$ or $a_Q=0$, therefore $l_q= l_{qmod}$. If both conditions are fulfilled then the two equations are independent of each other. However, if only one of the coupling parameters is zero, then the solution of the decoupled equation appears as a source term in the other side. In particular if $a_Q=0$, then 
	\begin{align}\label{r1pdec}
		\tau_q\dot{q} + q - l_q^2\partial_{xx}q + a_q\partial_x Q &=\lambda_{th}\partial_x \beta ,\\
		\label{r2pdec}
		\tau_Q\dot{Q} +Q-l_Q^2\partial_{xx}Q &=0.
	\end{align}
	
	Therefore, the solution of \eqref{r2pdec} -- a partial differential equation reminiscent of the Fourier one with an additional heat exchange source term --  provides a source term of the MCV type evolution equation of the heat flux according to \eqref{r1pdec}. This simple physical insight is not apparent from the {\it decoupled rate form}
	\begin{gather}\label{qrdec}
		\tau_Q\tau_q\ddot{q} + 	(\tau_q+\tau_Q)\dot{q} + q -
		(\tau_q l_Q^2 +\tau_Ql_q^2)\partial_{xx}\dot{q} -
		(l_Q^2 + l_q^2)\partial_{xx}q + l_Q^2l_q^2 \partial_{xxxx}q - \\ \nonumber
		-\lambda_{th} \left(\tau_Q\partial_{x}\dot\beta + \partial_x\beta-l_Q^2\partial_{xxx} \beta \right)=0,
	\end{gather}
neither from the simpler {\it decoupled hierarchical form}:
	\begin{gather}\label{qrdechier}
		(\tau_Qd_t+ 1- l_Q^2\partial_{xx})(\tau_q\dot{q} + q -l_q^2\partial_{xx}\dot{q} - \lambda_{th}\partial_x\beta)=0.
	\end{gather}
	
	The comparison of the differential operators of the hierarchical \eqref{qrdechier}, and the decoupled system forms \eqref{r1pdec}-\eqref{r2pdec} of heat conduction equations is remarkable regarding the interpretation of any hierarchical arrangements of the evolution equations. 
	
	It is straightforward to check that in both cases, since $l_q^2+l_Q^2 \geq 0$, the inequality \eqref{dstabcond2} is fulfilled. Therefore, the homogeneous equilibrium of decoupled heat conduction is naturally stable, the linear stability is the consequence of the Second Law of Thermodynamics. 
	
	\subsubsection{Fourier, Maxwell--Cattaneo--Vernotte and Guyer--Krumhansl: $a_q=a_Q=0$}
	
	The simplest heat conduction theories are obtained if \eqref{r1p} is independent of \eqref{r2p}. Then the Fourier heat conduction law emerges if both the relaxation time $\tau_q$ and the characteristic length $l_q$ are zero, too. In this case \eqref{r1p} is reduced to
	\begin{equation}\label{F}
		q= \lambda_{th} \partial_x\beta = -\lambda_F \partial_x T,
	\end{equation}	
	where $\lambda_F = \lambda_{th}/T^2$ is the Fourier heat conduction coefficient. 
	
	Also, if the relaxation time or the characteristic length is nonzero, one can recover the Maxwell--Cattaneo--Vernotte equation
	\begin{equation}\label{MCV}
		\tau_q\dot{q} + q +\lambda_F\partial_x T =0,
	\end{equation}
	and the Guyer--Krumhansl equation, respectively
	\begin{equation}\label{GK}
		\tau_q\dot{q} + q - l_q^2\partial_{xx} q +\lambda_F\partial_x T =0.
	\end{equation}
	Both \eqref{MCV} and \eqref{GK} are differential equations, and together with the energy balance \eqref{1} require proper initial and boundary conditions. Natural boundary conditions can be understood by second law analysis, as it was researched e.g. in \cite{BerVan17b,SzuEta20a,szucs24investigating}. 
	
	The stability polynomial of the Guyer--Krumhansl case is
	\begin{equation}\label{sGK}
		\tau_q\Gamma^2 +\left(1 + l_q^2k^2\right)\Gamma+\frac{\lambda_F}{\rho c_v}k^2 =0.
	\end{equation}
	Therefore, one can check directly that the linear stability of homogeneous equilibrium is natural, in the sense that follows from the thermodynamic conditions.

	\subsection{Extended Thermodynamics: $l_q=0$ and $l_Q=0$}

	In this case the second spatial derivatives in \eqref{r1p} and in \eqref{r2p} are zero, and therefore, the system becomes analogous to the one dimensional version of the 9 field equations of Rational Extended Thermodynamics (see \cite{MulRug98b} page 351, eq. (3.11)), and Extended Irreversible Thermodynamics, therefore can be compared directly to heat conduction theories that are compatible with kinetic theory. Let us remark that a derivation that is based on a particular microscopic interpretation (e.g. rarefied monoatomic or polyatomic gases) cannot be considered general. The obtained system of equations is referred to as {\em ballistic-conductive} in \cite{KovVan15a}. 
	
	The system of equations has the following form
	\begin{align}\label{r1pext}
		\tau_q\dot{q} + q + a\partial_x Q &=\lambda_{th}\partial_x \beta ,\\ \label{r2pext}
		\tau_Q\dot{Q} + Q + a\partial_x q &=0.
	\end{align}
	
	Here it was already considered that nonnegative entropy production requires the coupling coefficients to be equal:
	\begin{equation}
		a := a_q=a_Q,
	\end{equation}
	because the zero characteristic lengths $l_q=l_Q=0$ stem from the conditions $\lambda_{22}=0$ and $\kappa_{11}=0$ for the original thermodynamics coefficients, therefore the third inequalities of \eqref{stcond1} and \eqref{stcond2} demand that the conductivity matrix is antisymmetric. This is also observed by Rukolaine \cite{Ruk23a}.
	
	Then the \textit{rate equation} of the heat flux, \eqref{qr1}, for the Extended Thermodynamics case becomes
	\begin{gather}\label{qrET}
		\tau_Q\tau_q\ddot{q} + 	(\tau_q+\tau_Q)\dot{q} + q - a^2 \partial_{xx} q-
		\lambda_{th} \left(\tau_Q\partial_{x}\dot\beta + \partial_x\beta \right)=0.
	\end{gather}
	
	The corresponding stability polynomial follows as
		\begin{equation}\label{15c}
			0 =\tau_q\tau_A\Gamma^3 + 
			\left[\tau_q+ \tau_Q\right]\Gamma^2 +\left[1 + 
			\left(\tau_Q\alpha + a^2 \right)k^2\right]\Gamma + k^2\alpha.
		\end{equation}
	According to the Routh--Hurwitz criteria \eqref{17},
		\begin{equation}\label{17c}
		0  < a_1a_2-a_0a_3 = \left[\tau_Q^2\alpha 
		+\left(\tau_q+\tau_Q\right)\underline{a^2}\right]k^2
		+\left(\tau_q+\tau_Q\right).
	\end{equation}	
	
	The linear stability of homogeneous equilibrium of heat conduction in Extended Thermodynamics follows from thermodynamic conditions, because the inequality of stability condition, \eqref{dstabcond2} is reduced to $a^2 \geq 0$. 
	
	Materials that obey the above heat conduction equation have the interesting property, that the initial internal energy is locally accumulated, therefore it has a particular heat storage capability, independently  of the heat capacity of the material. This property is due to the hyperbolicity, the lack of diffusive terms, that is the second order spatial derivatives in the system of equations. The local  accumulation of internal energy may be considered unphysical as was observed recently by Rukolaine \cite{Ruk23a,Ruk24a}. It is also a characteristic property of the Jeffreys equation \cite{RukSam13a}.
	
	It is also remarkable that if the coupling coefficients $a_q$ and $a_Q$ are calculated from kinetic theory calculations then they are not necessarily equal. For example in \cite{MulRug98b}  $a_q = c^2 \tau_R$ and $a_Q = \frac{4}{15} \frac{\tau_R\tau_N}{\tau_R+\tau_N}$, where $c$ is the Debye speed, $\tau_R$ and $\tau_N$ are the relaxation times of the energy conserving resistive R-processes and energy and momentum conserving normal N-processes of the Callaway model (see also the comparative study in \cite{KovEta21a}). In the recent thermodynamic framework difference between $a_q$ and $a_Q$ is possible only if at least one of the characteristic lengths, $l_q$ or $l_Q$, are not zero. Then one of the diffusion terms, the second spatial derivatives of $q$ or $Q$ reappear in the transport form, \eqref{r1p}-\eqref{r2p}. However, those terms cannot appear in Extended Thermodynamics.

	\subsubsection{Burgers form:  $l_q=0$, $l_Q=0$ and $a=0$}
	
	If $a=0$ in \eqref{qrET}, then it looks like a Burgers equation
	\begin{gather}\label{qrB}
		\tau_Q\tau_q\ddot{q} + 	(\tau_q+\tau_Q)\dot{q} + q - 
		\lambda_{th} \left(\tau_Q\partial_{x}\dot\beta + \partial_x\beta \right)=0.
	\end{gather}
	
	However, a general Burgers equation with the reciprocal temperature looks like 
	\begin{gather}\label{genB}
		\nu_1\ddot{q} + \nu_2\dot{q} + q + \nu_3\partial_{x}\dot\beta -\lambda_{th}\partial_x\beta=0.
	\end{gather}
	with four independent coefficients $\nu_1,\ \nu_2,\ \nu_3$ and $\lambda_{th}$. In \eqref{qrB} the four coefficients are expressed by three parameters, $\tau_q$, $\tau_Q$ and $\lambda_{th}$, therefore the Burgers equation is degenerated. It is evident if it is rearranged in a hierarchical form, just like in \eqref{qrdechier}, as
	\begin{gather}\label{qrBhier}
		(\tau_Qd_t+ 1)(\tau_q\dot{q} + q - \lambda_{th}\partial_x\beta)=0,
	\end{gather}
	or remember that the system is decoupled and the rate equation of heat conduction is actually a  Maxwell--Cattaneo--Vernotte one.

	\subsubsection{Jeffreys equation: $l_q=l_Q=0$, $a=0$ and $\tau_q=0$}
	
	A general Jeffreys type equation in term of the reciprocal temperature $\beta$ appears if $\nu_1=0$ in the general Burgers equation \eqref{genB}, and can be given in the following form:
	\begin{gather}\label{genJ}
		\nu_2\dot{q} + q +\nu_3\partial_{x}\dot\beta -\lambda_{th}\partial_x\beta=0.
	\end{gather}
	
	In the nonequilibrium thermodynamic framework it is worth to look back to \eqref{qr}, and observe that the terms of the Jeffreys equation are nonzero if and only if $M\neq 0$, $\kappa_{22}\neq 0$ and $\lambda_{11}\neq 0$. All the other terms must be zero, therefore $m = 0$, $\lambda_{22}= 0$, $\kappa_{11} = 0$ and $\check \lambda = \check \kappa$. The last condition follows, because the coefficient of $\partial_{xx}q$ must be zero. This coefficient appears in the ultimate stability condition of the general equation,  in \eqref{dstabcond}. Considering the previous conditions $\lambda_{22}= 0$ and $\kappa_{11} = 0$ together with the least inequalities of \eqref{stcond1}  and \eqref{stcond2} follows that $\hat\lambda=0$ and $\hat\kappa =0$, the symmetric part of the coefficient matrices are zero. Therefore the last term of  \eqref{dstabcond} must also be zero.  

 Then one obtains the following form
    \begin{gather}\label{qrJorig}
		\lambda_{11}M\dot{q}+\lambda_{11}\kappa_{22}q-M\partial_x\dot{\beta}-\kappa_{22}\partial_x\beta=0.
	\end{gather}
 
       If it is expressed with physical coefficients, or equivalently, the relaxation time $\tau_q$ is zero in \eqref{qrB}, then it is written as
	\begin{gather}\label{qrJ}
		\tau_Q\dot{q} + q - \lambda_{th}\tau_Q\partial_{x}\dot\beta - \lambda_{th}\partial_x\beta=0.
	\end{gather}
	
	This looks like a Jeffreys type heat conduction equation. However, there are three coefficients and only two independent parameters. It is apparent both in the hierarchical form
	\begin{gather}\label{qrJhier}
		(\tau_Qd_t+ 1)(q - \lambda_{th}\partial_x\beta)=0,
	\end{gather}
	and the original decoupled system \eqref{r1pext}-\eqref{r2pext}, that it is a Fourier equation for the heat flux $q$.
 
	Therefore, neither the general Burgers nor the general Jeffreys equation can be derived with the recent conditions. That fact remained hidden in \cite{KovVan15a}. Also, the degenerate Burgers and Jeffreys equations are linearly stable, but thermodynamics of Extended Heat Conduction is silent about the stability of the general ones. The observed degeneracy of the evolution equations is unexpected.

	\subsection{Heat conduction without Fourier: $\lambda_{11}=0$}
	
	A particular family of general heat conduction equations does not contain Fourier type heat conduction as a special case. For example for the Green--Naghdi II equations  the heat flux part is simply missing, moreover, the dissipation is zero. Such a possibility is not apparent in the previous investigations because it is connected to singularities of the physical coefficients, \eqref{phcoeff1}. However, considering the original thermodynamic conductivity parameters one may recognize, that nothing excludes, that $\lambda_{11}$, the coefficient of the heat flux in \eqref{o2}, be zero. 
	
	Then the system of transport form equations \eqref{r1}-\eqref{r2} cannot be rewritten with some of the previous physical parameters, but it is worth introducing different ones
	\begin{align}\label{r1nF}
		\dot{q} -\xi_q \partial_{xx}q + A_q\partial_x Q - \kappa \partial_x \beta &= 0,\\
		\label{r2nF}
		\tau_Q\dot{Q} + Q- l^2_Q\partial_{xx}Q + a_Q\partial_x q &=0.
	\end{align} 
	
	Here the physical parameters in the second equation remained the same as before, but the following coefficients were introduced in \eqref{r1nF}:
	$$
	\xi_q := \frac{\kappa_{11}}{m}, \quad
	A_q:= \frac{\kappa_{12}-\check\lambda}{m}, \quad 
	\kappa := \frac{1}{m}.
	$$
	Nonnegative entropy production with $\lambda_{11}=0$ requires the related offdiagonal elements to have an opposite sign, therefore $\check\lambda =\lambda_{12} = - \lambda_{21}$.
	
	Now, the rate form can be written as
	\begin{gather}\label{qrnF}
		\tau_Q \ddot{q} + \dot{q} - (l_{Q}^2 + \tau_Q\xi_q)\partial_{xx}\dot{q} -
		(\xi_q + a_QA_q)\partial_{xx}q +
		\xi_ql^2_Q \partial_{xxxx}q = \\ \nonumber
		\kappa(\tau_Q\partial_{x}\dot\beta + \partial_x\beta-l_Q^2\partial_{xxx} \beta).
	\end{gather}
	
	The hierarchical form becomes 
	\begin{gather}\label{qrnFhier_gen}
		(\tau_Qd_t- l_Q^2\partial_{xx})(\dot{q} -\xi_q\partial_{xx}{q} - \kappa\partial_x\beta)+
		(\dot{q} -\check \xi_q\partial_{xx}{q} -\kappa\partial_x\beta)=0.
	\end{gather}
	Here $\check \xi_q =\xi_q + a_QA_q $. Previously in \eqref{qrdechier}, the hierarchical structure revealed two Guyer--Krumhansl equations with different characteristic lengths. Now, the heat flux term is missing and instead of Guyer--Krumhansl equations one can observe two equations in the form 
	$\dot{q} -\xi\partial_{xx}{q} -\kappa\partial_x\beta =0$. Those are second type Green--Naghdi equations when the $\xi$ parameter is zero \cite{GreNag91a}. 
	
	The inequality condition for the linear stability of the homogeneous equilibrium \eqref{dstabcond}, in our case is written as
	\begin{gather}\label{dstabcond_}
		\check\lambda^2 +K+ \check\lambda(\kappa_{21}-\kappa_{12}) = 
		\kappa_{11}\kappa_{22} - \kappa^2 + (\check\lambda-\check\kappa)^2\geq 0.
	\end{gather}
	Therefore, according to \eqref{stcond2} the linear stability of homogeneous equilibrium is warranted by thermodynamic conditions.

	\subsubsection{General Quintanilla: $\xi_q=0$ and $l_Q^2=0$}

	In this case \eqref{qrnF} simplifies to 
	\begin{gather}\label{qrnFE}
		\tau_Q \ddot{q} + \dot{q} - a_QA_q\partial_{xx}q -\kappa(\tau_Q\partial_{x}\dot\beta + \partial_x\beta)=0.
	\end{gather}
	
	If the system is decoupled then $a_Q=0$ and \eqref{r1nF} is reduced to the Quintanilla equation \cite{Qui19a},
	\begin{gather}\label{qrnFQ}
		\tau_Q \ddot{q} + \dot{q} -\kappa(\tau_Q\partial_{x}\dot\beta + \partial_x\beta)=0.
	\end{gather}
	This is a particular form, like in the case of Jeffreys equation, with two independent coefficients only. The hierarchical rearrangement also reflects the decoupling
	\begin{gather}\label{qrnFhier}
		(\tau_Q d_t + 1)(\dot{q} +\kappa\partial_x\beta)=0.
	\end{gather}
	
	Based on the linear stability analysis calculations in Section \ref{sec_3}, the stability polynomial of \eqref{qrnFQ} is the following:
	\begin{equation}
		\tau_Q \Gamma^3 + \Gamma^2 + n\tau_Q \Gamma + n = 0,
	\end{equation}
	with the notation $n := k^2\kappa/\rho c_v$.
	
	Since the coefficients of the polynomial are nonnegative, only the extra inequality \eqref{17} would be restrictive. However, it is trivially fulfilled, indicating, that the homogeneous equilibrium of the Quintanilla equation is marginally stable. This means, that according to the applied Routh--Hurwitz criteria from Section \ref{sec_3}, eigenvalues of the Jacobian matrix do have real parts.

	\section{Summary and discussion}
		
\begin{figure}[h!]
		\centering
		\includegraphics[width=0.8\textwidth]{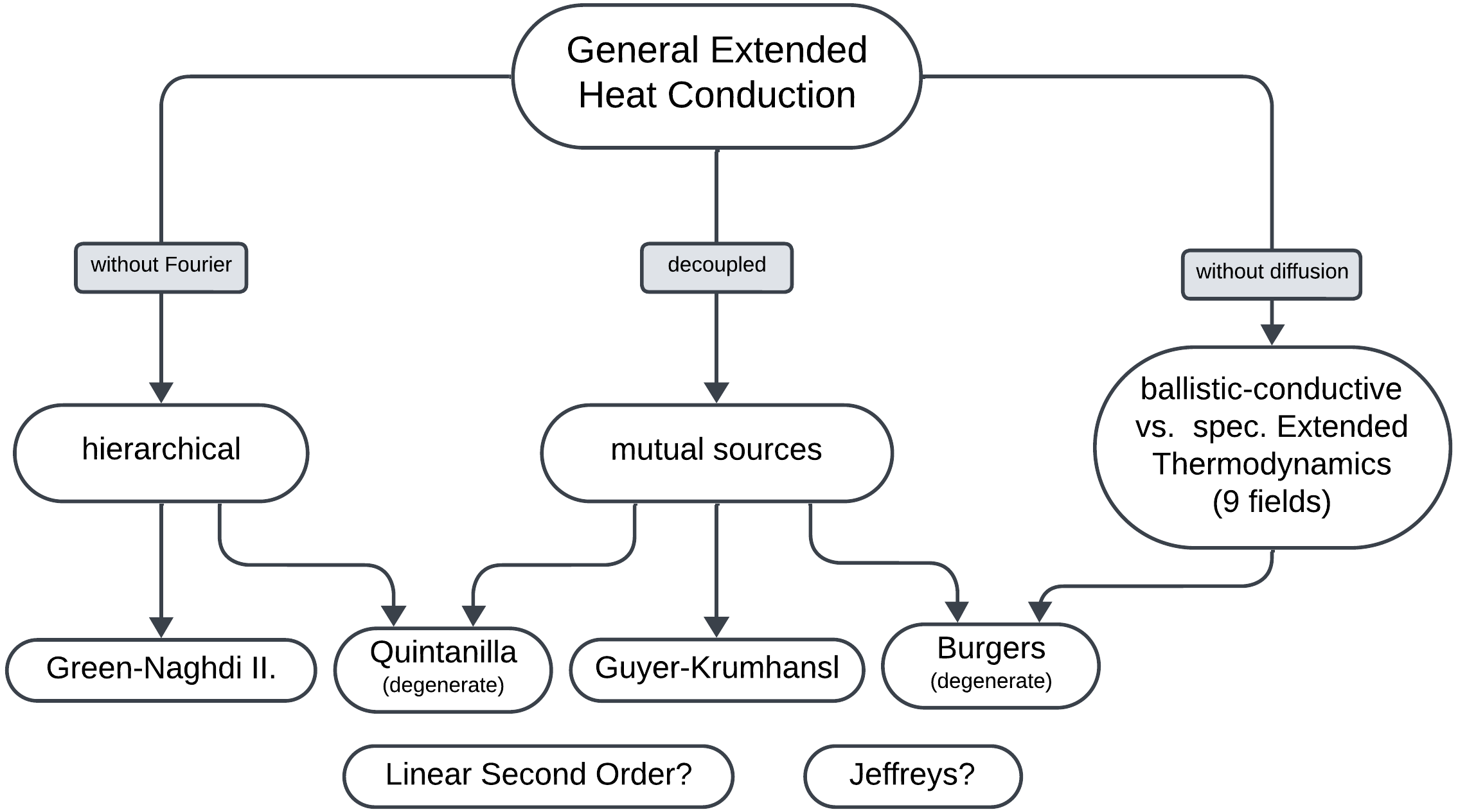}
		\caption{The relation of the structure of the evolution equations of the various heat conduction theories in one spatial dimension. Several know heat conduction theories are related through nonequilibrium thermodynamics. Also there are some theories and concepts in heat conduction modelling, that are not incorporated in this framework, like the nondegenerate Jeffreys type equation. \label{fig:thd_spechc}}
	\end{figure}

	In this paper a thermodynamic compatibility and stability analysis of third order non-relativistic Extended Theories of Heat Conduction have been developed. Several known special cases were investigated in the light of a general approach of Nonequilibrium Thermodynamics with Internal Variables (NeTIV), and the general framework provided a classification of several known heat conduction models and theories, as it is shown in Figure \ref{fig:thd_spechc}. In NeTIV, the deviation from thermodynamic equilibrium is characterised by additional fields, called internal variables. The evolution equation of the internal variables is derived taking into account the requirements of the Second Law of Thermodynamics. 
	
	Furthermore, the following modeling assumptions are necessary for compatibility with the moment series expansion of kinetic theory and therefore are characteristic in both Extended Irreversible Thermodynamics and Rational Extended Thermodynamics: 
	\begin{enumerate}
		\item Increasing tensorial order of the basic fields. Therefore, two internal variables were introduced in this work: a vectorial and a tensorial one. 
		\item The vectorial internal variable is the heat flux. 
		\item Both the vectorial and the tensorial internal variables vanish in thermodynamic equilibrium. 
	\end{enumerate}
	
	As a consequence of the third assumption the entropy is a quadratic function of the internal variables \cite{MacOns53a,Gya77a}, and the entropy flux is proportional to the internal variables \cite{Nyi91a1}. These properties do not require a particular microstructure, like rarefied gases, they are general. Therefore, the universality of the consequent heat conduction theories, like Fourier theory, follows. 
	
	The investigations are restricted to isotropic materials and to one spatial dimension. By eliminating the tensorial internal variable, the {\em rate equations of the heat flux} were obtained, so that the deviation from thermodynamic equilibrium was transformed into a deviation from the Fourier law, the usual framework of non-Fourier heat conduction theories. 
 \vspace{2cm}
	

	There are two main implications of our research: 
	\begin{itemize}
		\item The thermodynamic conditions -- concave entropy and non-negative entropy production -- ensure the linear stability of homogeneous equilibrium in most cases. However, the possible violation appears on the most general level. The special cases, like theories of Extended Thermodynamics, are stable, because the instability is connected to nonzero characteristic lengths of the diffusion of the heat flux $l_q$, and the diffusion of the flux of the heat flux $l_Q$, and those terms are absent in Extended Irreversible Thermodymamics and Rational Irreversible Thermodynamics. 		
		\item One can obtain evolution equations of many known heat conduction theories in the presented uniform theoretical framework of nonequilibrium thermodynamics. However, in several important cases the derived equations are degenerate. For example, in the derived Jeffreys-type equation the 3 material coefficients were not independent. A similar situation was observed for the 9-field theory of Extended Thermodynamics, for the Burgers equation and also for the Quintanilla equation. 
	\end{itemize}
	
	Regarding the first consequence, it should be noted that in the absence of a microscopic background, we have not investigated Onsager or Casimir type symmetries. The results of our work, in particular the stability condition, are independent of any kind of reciprocity relations. However, if one considers that the heat flux $q$ is odd by time reversal and the second order tensor $Q$ is even (that follows e.g. from kinetic theory interpretation), then Casimir symmetry is valid for the cross-coefficients, hence $\lambda_{12} = -\lambda_{21}$ and $\kappa_{12} = -\kappa_{21}$. Therefore, the stability condition \eqref{dstabcond} is satisfied and the linear stability of the homogeneous equilibrium follows from the second law. However, this argument is based on the concept of macroscopic reversibility \cite{Mei75a}, and requires further analysis, in particular considering some new developments in this respect, like \cite{PavEta14a,Gav23a}.

    In the introduction we argued that FDS (fundamental dynamic stability) is expected by pure thermodynamic conditions because thermodynamics is a stability theory. Regarding the dynamic stability of other equilibria, like the stationary states with constant fluxes, we expect stability and instability conditions. It is easy to see that thermodynamic stability of the $(T,q,Q)=(T_0,q_0,0)$ steady state leads to a condition: above the maximal heat flux, the entropy becomes convex, and instability appears. A detailed stability analysis of steady states in non-Fourier heat conduction is a yet unexplored area of research.
 
	With regard to the second consequence, it can be noted that the reason why the most general equations of heat conduction could not be obtained is apparently due to the special choice of the basic fields. In fact, the deviation from equilibrium was characterised by a vectorial internal variable with a fixed physical interpretation: it was the heat flux according to our second assumption. If the heat flux is not a basic field, but a constitutive quantity, in the sense of Fourier heat conduction, then one can expect more general heat conduction theories, with independent coefficients. For example, in case of a single vectorial internal variable one can obtain the Jeffreys equation without restrictions \cite{VanFul12a,CiaRes19a,Kov24a}. According to this conjecture, removing the restrictive interpretation of the vectorial internal variable being the heat flux, there is a hope that one could understand the relation of existing heat conduction theories within the single, uniform and constructive framework of NeTIV. 
	
	A final remark seems to be suitable at the end. The form of the equations in itself is not the whole theory. First of all, our analysis is restricted to one spatial dimension, the various named theories are distinctly different in a complete spacetime compatible framework. That complete framework defines the physical parameters, the mathematical problems and also determines the experimental approach. The physical parameters are comprehended differently in kinetic theory motivated approaches and in nonequilibrium thermodynamics. The boundary conditions and the structure of the equations determines the well or ill posedness of the related mathematical  problems. Finally, the solution of the equations and the notions of the theory (e.g. universal or microstructure dependent) guide the design of the experiments. Therefore, the outlined connections of Figure \ref{fig:thd_spechc} should be treated as guiding analogies.

 \section{Acknowledgements}

The authors thank R\'obert Kov\'acs and M\'aty\'as Sz\"ucs for valuable discussions. The research reported in this paper has been supported by the NRDI Fund (No. TKP2020 NC, Grant No. BME-NCS) based on the charter of bolster issued by the NRDI Office under the auspices of the Ministry for Innovation and Technology.
 
	\newpage
	\bibliographystyle{unsrt}

\end{document}